\documentclass{camera}
\usepackage{graphicx}  % uncomment this if your want to insert figures
\usepackage{amssymb,amsmath,amsthm,bm}

\begin{document}

%%%%%%%%%%%%%%%%%%%%%%%%%%%%%%%%%%%%%%%%%%%%%%%%%%%%%%%%
% The title, only the first letter capitalized; if you want to split it in
% two or more lines, put a \\ macro at each line break
% example:
%   \title{Title: first line\\ second line}
%
\title{How do neutrinos oscillate in moving and accelerated matter?}

%%%%%%%%%%%%%%%%%%%%%%%%%%%%%%%%%%%%%%%%%%%%%%%%%%%%%%%%
% The author(s), separated by commas; do not put a
% comma before the last author, use instead the \and
% macro which produces a normal ``and'' in the
% caps/small caps context
%
\author{Alexander Studenikin$^{1,2}$ \and Ilya Tokarev$^1$}

%%%%%%%%%%%%%%%%%%%%%%%%%%%%%%%%%%%%%%%%%%%%%%%%%%%%%%%%
%
\organization{$^1$Department of Theoretical Physics, Faculty of Physics, Moscow State University, Moscow 119991, Russia\\
$^2$Joint Institute for Nuclear Research, Dubna 141980, Moscow Region, Russia}

\maketitle

\begin{abstract}
Neutrino flavour oscillations in a nonuniformly moving matter are considered. The neutrino oscillation resonance condition in presence of matter, in the most general case when matter is moving with acceleration, is derived for the first time. We predict that the effect of matter acceleration can have significant influence on neutrino oscillations pattern in different astrophysical environments.
\end{abstract}

%%%%%%%%%%%%%%%%%%%%%%%%%%%%%%%%%%%%%%%%%%%%%%%%%%%%%%%%
% Write the text starting from here and using the usual
% LaTeX commands.

In this short note we generalize our previous studies \cite{GrigLobStud2002} on neutrino flavour oscillations in an uniformly moving matter to the case when matter moves with acceleration.

Consider two flavour (electron and muon) neutrino oscillations in nonuniformly moving matter. The effective Lagrangian of neutrino interactions with background matter can be expressed in the form
\begin{equation}
\label{L}
\mathcal{L}_{eff}=-\bar{\nu}\left(\gamma_{\mu}f^{\mu}\frac{1+\gamma^5}{2}\right)\nu, \quad
f^{\mu}=\frac{G_Fn\gamma}{\sqrt{2}}(1,\textbf{\textit{v}}),
\end{equation}
where $G_F$ is the Fermi constant, $\bm v$ is the speed of matter, $\gamma=(1-\bm v^2)^{-\frac12}$ and $n=n_0$ and $n=-n_0$ for the electron and muon neutrino correspondingly. Here $n_0$ is the electron number density in the reference frame for which the total speed of matter is zero. Note that we consider the matter composed of neutrons, protons and electrons. Obviously, neutrons and protons do not influence the oscillations.

For the effective Hamiltonian of neutrino flavour states evolution we obtain
\begin{equation}
\label{H}
H=\sigma_1\frac{\delta m^2\sin2\theta}{4E}-
\sigma_3\left(\frac{\delta m^2\cos2\theta}{4E}-
\frac{G_Fn_0}{\sqrt{2}}\frac{1-v(x)\cos\phi}{\sqrt{1-v^2(x)}}\right),
\end{equation}
where $\delta m^2=m_2^2-m_1^2$, $\theta$ is the vacuum mixing angle, $E$ is the neutrino energy and $\phi$ is the angle between the speed of matter and the direction of neutrino propagation (it is supposed that the neutrino is propagating along $x$ direction). In the adiabatic approximation the oscillation probability has the usual form $P_{\nu_e\rightarrow\nu_{\mu}}(x)=\sin^22\theta_{eff}\sin^2\frac{\pi x}{L_{eff}}$, where the effective mixing angle $\theta_{eff}$ and oscillation length $L_{eff}$ are determined by the elements $H_{ij}$ of the evolution Hamiltonian~(\ref{H}).

The straightforward calculations yields the neutrino flavour oscillations resonance condition~\cite{Wolf1978,MikhSmirn1985}
\begin{equation}
\label{res_cond}
\frac{\delta m^2}{2E}\cos2\theta=\sqrt{2}G_Fn_e(x), \quad \text{where} \quad
n_e(x)=\frac{n_0}{x}\int_0^x\frac{1-v(x')\cos\phi}{\sqrt{1-v^2(x')}}dx'.
\end{equation}
In the considered case the neutrino oscillations probability gets its maximum value in a set of points $x_k$ that are the solutions of Eq.~(\ref{res_cond}). Note that in general case Eq.~(\ref{res_cond}) is not linear in respect to $x$. Obviously, that in case of monotonic $x$ dependence of the density $n_e$ there can be only one resonance point.

Consider matter motion with a constant acceleration $a$. Then the matter velocity is given by $v(x)=\frac{v_0\gamma_0+ax}{\sqrt{1+(v_0\gamma_0+ax)^2}}$ where $v_0$ is the initial matter speed. In this case one can obtain the effective electron matter density in the form
\begin{equation}
\label{ne_v}
n_e=\frac{n_0}{2}
\left(\frac{\gamma V-\gamma_0 V_0+\ln\frac{\gamma+V}{\gamma_0+V_0}}{V-V_0}-
(V+V_0)\cos\phi\right), \quad V\equiv\gamma v.
\end{equation}

\begin{figure}[!]
\includegraphics[scale=0.21]{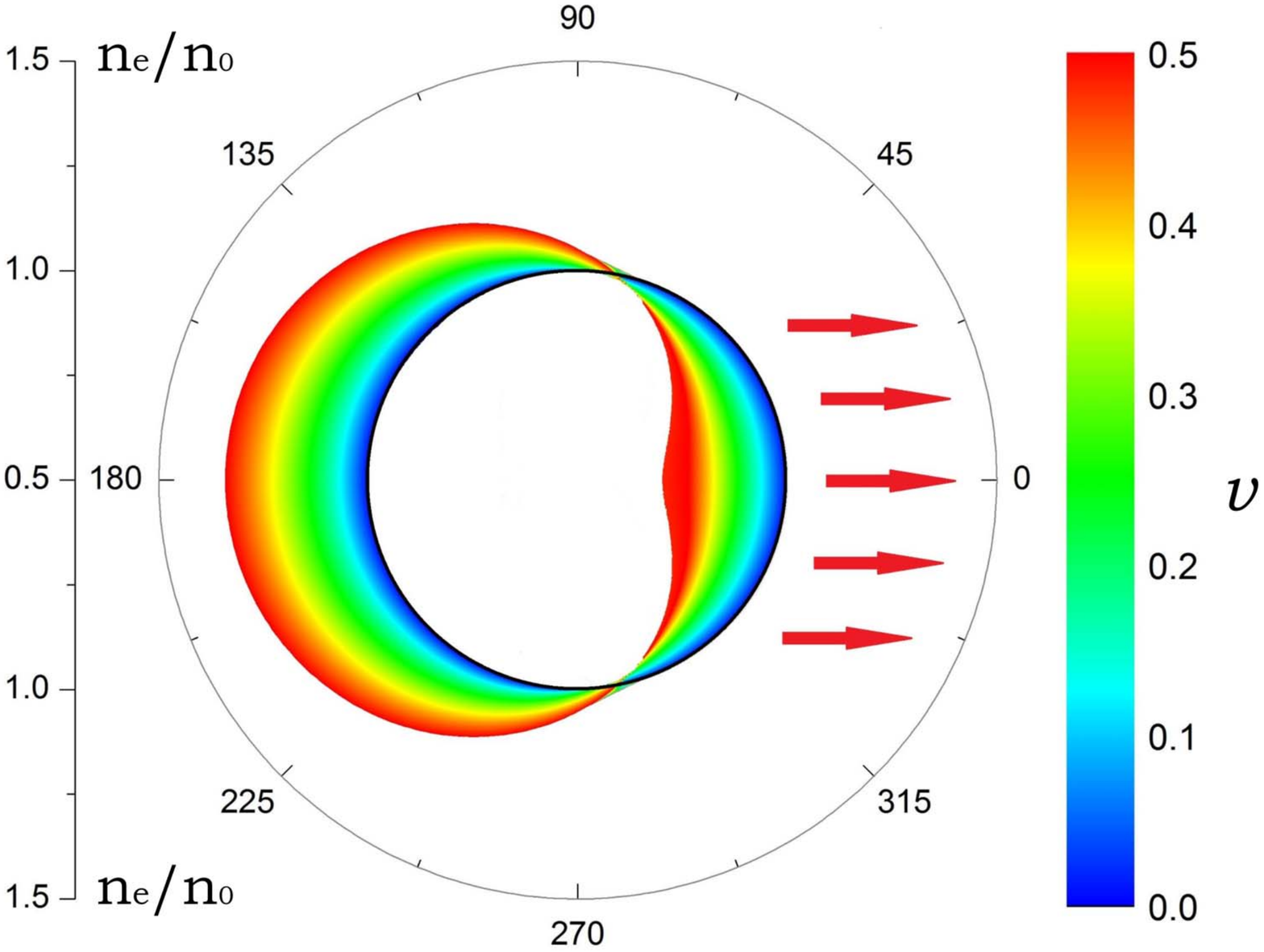}
\includegraphics[scale=0.21]{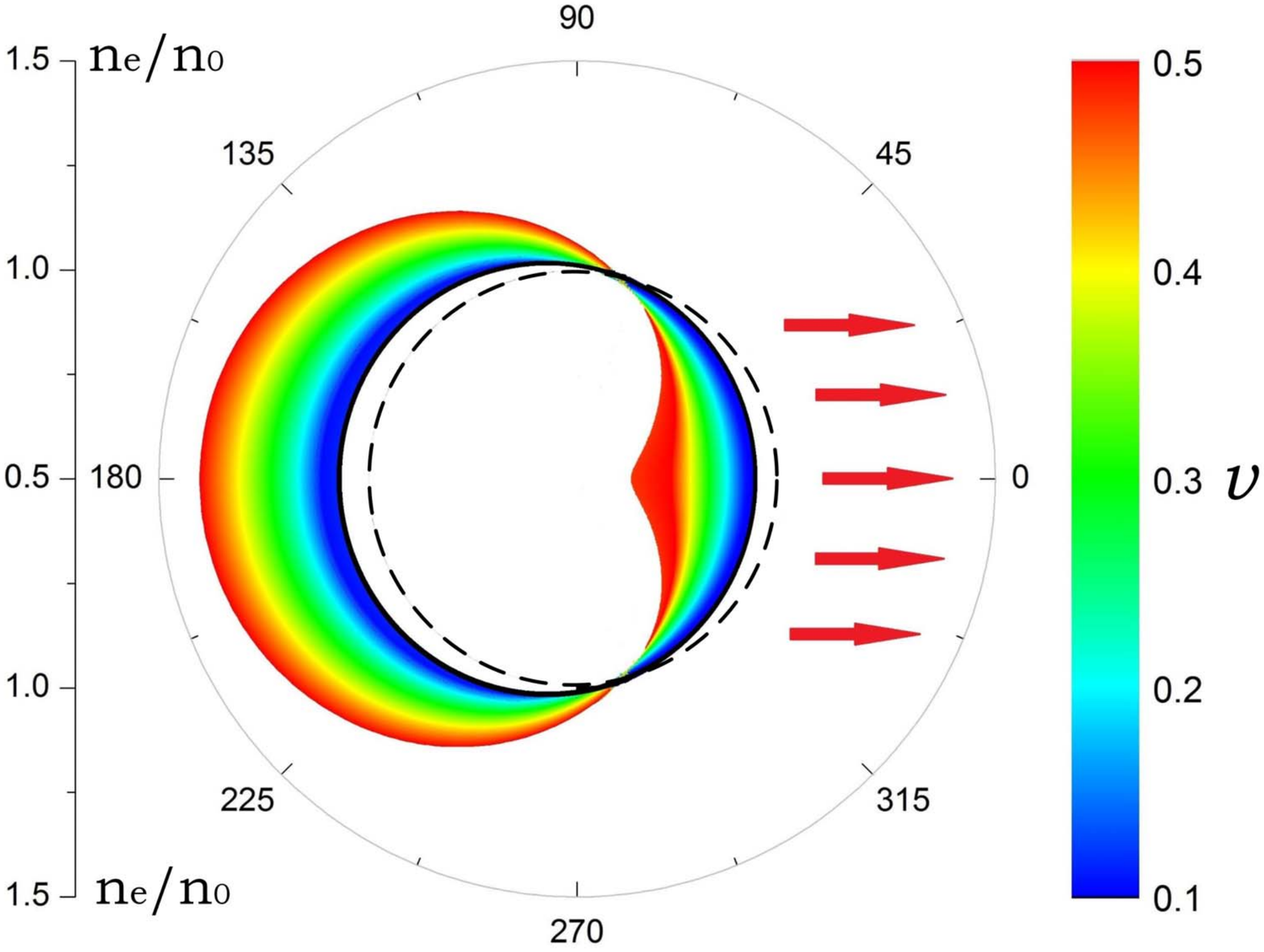}
\caption{The figure illustrates distribution
of the effective electron density in moving matter (normalized on the static value) as a function of
matter speed $v$ and the angle of propagation $\phi$. The red arrows defines the direction of
a neutrino flux. The left and right plots corresponds to $v_0=0$ and $v_0=0,1$ accordingly.}
\label{fig01} % optional figure label, must be unique
\end{figure}

In Fig.~\ref{fig01} the function $n_e=n_e(v(x),\phi)$ is plotted. It follows that $n_e$ significantly increases in case of matter motion against neutrino propagation ($\phi=\pi$) and almost vanishes in the opposite case ($\phi=0$). In the nonrelativistic and ultrarelativistic limits of matter motion Eq.~(\ref{ne_v}) can be simplified as follows
\begin{equation}
\label{ne_ax}
\frac{n_e}{n_0}=
\left\{
\begin{aligned}
&(4ax)^{-1},                &(ax\gg1, \phi=0),\\
&ax\sin^2\frac{\phi}{2},    &(ax\gg1, \phi\neq0),\\
&1-\frac{ax}{2}\cos\phi,    &(ax\ll1).
\end{aligned}
\right.
\end{equation}
In the corresponding evaluation it is supposed that $v_0=0$. When nonrelativistic matter is moving with a constant acceleration the corresponding shift of the electron number density is given by $\delta n(x)=\frac12 n_0\delta v(x)\cos\phi$, where $\delta n(x)=n_e(x)-n_0$ and $\delta v(x)=v(x)-v_0$. For instance, during a supernova core-collapse $\delta v(x)$ is up to 0,2 that yields an order of $10\%$ shift to the effective resonance number density $n_e$. It would be interesting to consider possibilities of observing this effect.

\section*{Acknowledgments}

This work has been supported by RFBR grants 14-02-00914, 14-02-31816.

%%%%%%%%%%%%%%%%%%%%%%%%%%%%%%%%%%%%%%%%%%%%%%%%%%%%%%%%
% End of the paper
%

\begin{thebibliography}{99}

\bibitem{GrigLobStud2002}
Grigoriev A., Lobanov A., Studenikin A., \textit{Phys.Lett.}, \textbf{B 535} (2002) 187.

\bibitem{Wolf1978}
Wolfenstein L., {\it Phys.Rev.} \textbf{D 17} (1978) 2369.

\bibitem{MikhSmirn1985}
Mikheyev S. and Smirnov A., {\it Sov.J.Nucl.Phys.}, \textbf{42} (1985) 913.

%\bibitem{Pal1992}
%Pal P., \textit{Int. J. Mod. Phys.} \textbf{A 7} (1992) 5387.

%\bibitem{LobStud2001}
%Lobanov A. and Studenikin A., {\it Phys. Lett.}, \textbf{B 515} (2001) 94.

%\bibitem{LikhStud1995}
%Likhachev G. and Studenikin A., \textit{Hadronic J.}, \textbf{18} (1995) 1.

\end{thebibliography}
\end{document}